\documentclass[10pt,journal,compsoc]{IEEEtran}
\ifCLASSOPTIONcompsoc
  \usepackage[nocompress]{cite}
\else
  \usepackage{cite}
\fi
\ifCLASSINFOpdf
\else
\fi
\usepackage{amsmath}
\usepackage{algorithm,algorithmic}
\usepackage{lineno,hyperref}
\usepackage{xcolor}
\usepackage{adjustbox}
\usepackage{blindtext}
\usepackage{titlesec}
\usepackage{graphicx}
\usepackage{caption}
\usepackage{subcaption}
\usepackage{lipsum}
\usepackage{color,soul}
\usepackage{array}
\usepackage{float}
\usepackage{comment}
\usepackage{titlesec}

\usepackage{multirow}
\usepackage{dblfloatfix}

\usepackage{tikz}
\def\checkmark{\tikz\fill[scale=0.5](0,.35) -- (.25,0) -- (1,.7) -- (.25,.22) -- cycle;}

\begin{document}
%
\title{Human Behavior-based Personalized Meal Recommendation and Menu Planning Social System}

%
%

\author{Tanvir~Islam, Anika~Rahman~Joyita, Md.~Golam~Rabiul~Alam,~\IEEEmembership{Member,~IEEE}, Mohammad Mehedi Hassan,~\IEEEmembership{Senior Member,~IEEE}, Md. Rafiul Hassan~\IEEEmembership{Senior Member,~IEEE}, and Raffaele Gravina
\IEEEcompsocitemizethanks{\IEEEcompsocthanksitem T.~Islam, A. R.~Joyita, M. G. R.~Alam are with the Department of Computer Science and Engineering, BRAC University, Dhaka,Bangladesh, 1212 (e-mail: tanvir.islam1@g.bracu.ac.bd, anika.rahman.joyita@g.bracu.ac.bd and rabiul.alam@bracu.ac.bd)

\IEEEcompsocthanksitem M. M. Hassan is with Department of Information Systems, College of Computer and Information Sciences, King Saud University, Riyadh, Saudi Arabia (e-mail: mmhassan@ksu.edu.sa)

\IEEEcompsocthanksitem M. R. Hassan is with the College of Arts \& Sciences, University of Maine at Presque Isle, ME04769, USA (e-mail: md.hassan@maine.edu)

\IEEEcompsocthanksitem R. Gravina is with the Department of Informatics, Modeling, Electronics, and Systems, University of Calabria, Italy (e-mail: r.gravina@dimes.unical.it).
}


}

%
%

\markboth{IEEE TRANSACTIONS ON COMPUTATIONAL SOCIAL SYSTEMS}%
{Shell \MakeLowercase{\textit{et al.}}: Bare Demo of IEEEtran.cls for Computer Society Journals}
%



\IEEEtitleabstractindextext{%
\begin{abstract}
The traditional dietary recommendation systems are basically nutrition or health-aware where the human feelings on food are ignored. Human affects vary when it comes to food cravings, and not all foods are appealing in all moods. It takes a lot of effort to learn people’s food preferences and make recommendations based on their affects and nutrition. A questionnaire-based and preference-aware meal recommendation system can be a solution. However, automated recognition of social affects on different foods and planning the menu considering nutritional demand and social-affect has some significant benefits over the questionnaire-based and preference-aware meal recommendations. A patient with severe illness, a person in a coma, or patients with \textit{locked-in syndrome} and \textit{amyotrophic lateral sclerosis (ALS)} cannot express their meal preferences. Therefore, the proposed framework includes a social-affective computing module to recognize the affects of different meals where the person's affect is detected using electroencephalography signals. EEG  allows to capture the brain signals and analyze them to anticipate affective state toward a food. In this study, we have used a 14-channel wireless Emotiv Epoc+ to measure affectivity for different food items. A hierarchical ensemble method is applied to predict affectivity upon multiple feature extraction methods and TOPSIS (Technique for Order of Preference by Similarity to Ideal Solution) is used to generate a food list based on the predicted affectivity. In addition to the meal recommendation, an automated menu planning approach is also proposed considering a person’s energy intake requirement, affectivity, and nutritional values of the different menus. The bin-packing algorithm is used for the personalized menu planning of breakfast, lunch, dinner, and snacks. The experimental findings reveal that the suggested affective computing, meal recommendation, and menu planning algorithms perform well across a variety of assessment parameters.
\end{abstract}

}

\maketitle
\IEEEdisplaynontitleabstractindextext

\IEEEpeerreviewmaketitle

\IEEEraisesectionheading{\section{Introduction}\label{sec:introduction}}

\IEEEPARstart{H}{uman} eating habits and preferences are difficult to anticipate because of diversity in food and human choices. However, menu planning and food recommendation systems are necessary for enabling a healthy lifestyle and well-being~ \cite{trattner2017investigating}. A menu planning and food recommendation system should consider human feelings on a diet along with the nutritional value of the food menu~\cite {harvey2012learning}. Even though a food recommender system recommends the most healthy diet, people will follow it only when the recommended
food appeals to the user’s taste buds~\cite {ge2015health}.

 Patients affected with \textit{locked-in syndrome} go under complete paralysis \cite{article} and unable to express preferred meals. The disease \textit{Amyotrophic lateral sclerosis (ALS)} affects nerves of the brain and gets a person towards paralysis, and loss of motor function \cite{KIERNAN2011942}. People affected by these types of diseases can not express their preference for foods. In this regard, neural signature-based food recommendations could be a more potent solution. Unfortunately, none of the state-of-the-art food recommendation systems consider the human affect on different foods and do not even use neural signatures for computing food affect. 
 
  The role of affective behaviour in recommender systems and decision-making is  presented in~ \cite {tkalvcivc2016emotions}\cite {picard2004affective}. Although quantifying human affects are challenging, and many researchers successfully classify human affects through “Affective computing”~\cite {picard1999affective} from electroencephalography signals (EEG)~\cite{khan2022cnn} and facial expressions~\cite{siddiqi2016novel}. This research also used EEG signals to determine human affects on different foods. Moreover, this research will answer the following research questions: 

\begin{itemize}
    \item How human affects on different foods can be extracted by analyzing the brain's spontaneous electrical activity?
    
    \item How a personalized food menu can be recommended considering the subject's calorie requirements, affective preferences, and nutrition value of the food menu?
\end{itemize}

According to~\cite{ganley1989emotion}, in certain emotional states, obese people intake more food than a person with normal weight. Therefore, the primary objective of this research is to develop a system that can assist users by recommending meals based on the person’s emotions. By incorporating the real-time emotion detection process, the proposed system will prepare meal plans that are compatible with the users' feelings. In addition to that, a full-day meal planning system has also been proposed that incorporates both emotion and nutritional data. 

 Table~\ref{workSummaryTable} shows that most state-of-the-art meal recommendation systems \cite{agapito2018dietos} \cite{espin2016nutrition} \cite{ribeiro2017souschef}\cite{toledo2019food} \cite {yang2017yum} are 
nutritional information and preference based. However, the important affects e.g., feelings and excitement of humans on different foods were not considered in food recommendations. Moreover, none of the recommender systems had considered neural signature-based methods for the automated extraction of human food preferences. Therefore, this research proposed Affective computing-based Meal Recommendation and menu Planning social system (AMRP). 
 

 



\begin{table}[H]
\Large
\caption{A Comparative Study of the Proposed AMRP Scheme with the  State-of-the-art Food Recommender Systems in Terms of Considered Factors}
\begin{adjustbox}{width=\columnwidth,center}
\fontsize{9}{11}\selectfont
$\begin{array}{|c||l|l|l|l|}

\hline \text { Research Work } & 
 
\begin{array}{l}
\text { Nutrition } \\
\text { aware } 
\end{array} & 
\begin{array}{l}
\text { Preference- } \\
\text { aware }
\end{array} & 
\begin{array}{l}
\text { Diet - } \\
\text { planning }
\end{array} & 
\begin{array}{l}
\text {Affective}\\
\text{state}\\
\text{aware}
\end{array} \\
\hline 
\text { Agapito et al. \cite{agapito2018dietos} }   & \checkmark & & &\\
\text { Espin et al. \cite{espin2016nutrition} } & \checkmark & & &\\
\text { Mata et al. \cite{MATA2018837}} &\checkmark & &\checkmark & \\
\text { Taweel et al. \cite{taweel2016service}} &\checkmark & & \checkmark&\\
\text { Bianchini et al. \cite{BIANCHINI201764}}  &\checkmark & \checkmark & \checkmark&\\
\text { Cioara et al. \cite{CIOARA2018368} }   &\checkmark & & & \\

\text { Elsweiler  et al. \cite{harvey2013you} }   &\checkmark & \checkmark & &\\

\text { Geleijnse  et al. \cite{geleijnse2011personalized} }   & & \checkmark & &\\

\text { Hernández-OcaÃśa et al. \cite{hernandez2018bacterial} }   &\checkmark & & & \\
\text { Syahputra et al. \cite{syahputra2017scheduling} }   &\checkmark & &  & \\
\text { Rehman et al. \cite{rehman2017ul} }   &\checkmark & &  & \\
\text { Ntalaperas et al. \cite{ntalaperas2015disys} }   & \checkmark &\checkmark & & \\

\text { Sevensson  et al. \cite{svensson2005designing} }   & & \checkmark & &\\
\text { Ribeiro et al. \cite{ribeiro2017souschef} }   & \checkmark &\checkmark & &\\
\text { Nag et al. \cite{nagLive} }   & \checkmark & & &\\
\text { Yang et al. \cite{yang2017yum} }   & \checkmark & \checkmark& & \\
\text { Ge et al. \cite{ge2015health} }  & & \checkmark& & \\
\text {{{Shandilya et al. \cite{shandilya2022mature})}} }  &\checkmark & \checkmark & \checkmark &  \\
\text {\textbf{Proposed method (AMRP)} }  &\checkmark & \checkmark & \checkmark & \checkmark \\
\hline
\end{array}$
\end{adjustbox}

\label{workSummaryTable}
\end{table}

The key contributions of this research are as follows: 
\begin{itemize}
      \item A meal recommendation and diet planning system has been proposed for an individual considering human affect toward that food, the nutritional value of the food, and the calorie requirement of the individual. 
      \item To quantify food affection, EEG signals of individuals were collected while presenting images of different food menus as stimuli. The self-assessment of likeness, valence, and arousal to the food menu were also collected using a Likert scale to determine the ground truth. 
    \item Three feature extraction methods: Short Time Fourier Transform (STFT), Discrete Wavelet Transform (DWT), and Hilbert-Huang Transform (HHT) were used to extract discriminative features from the EEG signals. In order to classify likeness, valence, and arousal, a hierarchical ensemble learning method was used.
    \item Based on the predicted affectivity, the preference, feelings, and excitement-aware food recommendation system has been proposed using the Technique for Order of Preference by Similarity to Ideal Solution (TOPSIS) method.
    \item A menu planning system has been proposed using Bin Packing algorithm considering affective state, calorie requirements along with the nutritional value of different food items.
   
\end{itemize}

The rest of the paper is presented as follows. Section II discussed the state-of-the-art food recommendation and diet recommendation systems. The proposed AMRP system is described in Section III.   Section IV discussed the implementation, results, and performance of the proposed model. Finally, Section V concludes the paper with some future directions. 

\begin{figure*}

 \centering

  \includegraphics[scale=0.5]{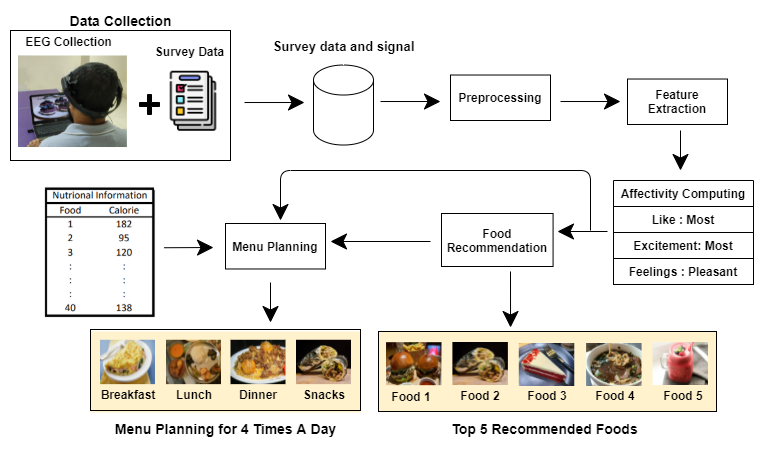}

  \caption{Top Level  Overview of the Affective Behavior-based Personalized Meal Recommendation and Menu Planning (AMRP) Social System}

  \label{architecture}

\end{figure*}

\section{Related Works}

The recommender system is a well-studied research domain and there are numerous recommendation engines for product recommendations, book recommendations, and movie or content recommendations. However, meal or food recommendation is different than the traditional recommendation engine because of the diversity of foods, people's choices, human feelings on foods, food nutrition, and individuals' diet requirements. A few pieces of research have been done on food recommendations where dietary preferences were learned through their online activity including ratings\cite{elsweiler2015towards}\cite{harvey2013you}\cite{forbes2011content}\cite{freyne2010intelligent}, previous recipe selections \cite{geleijnse2011personalized}\cite{svensson2005designing} as well as browsing history \cite{ueda2014recipe}\cite{van2011deriving}.

 Ribeiro et al., presented SousChef, a mobile supper recommender framework to assist older adults by providing a nutrition companion to guide them in making wise judgments about nourishment administration and healthy eating practices\cite{ribeiro2017souschef}. Their target demographic was senior adults who typically struggle to make good judgments about meal preparation, healthy eating, and grocery shopping. Studies also suggest that many older adults neglect nutrition and are more inclined to do so if they happen to live alone \cite{ramic2011effect}. The meal suggestion system was created to provide tailored nutritional programs based on the user's knowledge, which included personal preferences, activity level, and anthropocentric measurements (weight and height).

Jiang et al., developed "Market2Dish," a personalized health-aware food recommendation system that aids the user in developing a healthy eating habit \cite{jiang2019market2dish}. In this method, the ingredients are first studied by identifying them in micro-videos obtained from the market, then the user's health is observed by gathering data from their social media accounts, and finally customised healthy foods are recommended. Moreover. user's pathological report based diet recommendation system is proposed in Diet-Right\cite{rehman2017ul}. 

Gao et al., proposed the Progressive Attention-based Nourishment Suggestion (HAFR) system, which is capable of 1) capturing the collaborative sifting impact, such as what comparative clients tend to eat; 2) inferring a user's preference at the ingredient level; and 3) learning user preference from the recipe's visual images \cite{gao2019hierarchical}. Food proposals were characterized as an interactive media assignment in this work by taking into account user-recipe intuitive, culinary photos, and food ingredients. This paper suggested a system for inducing users' preferences for meal suggestions based on recipes.

Hsiao and Chang, a multi-objective optimization based location-aware diet recommendation system is proposed in SmartDiet\cite{hsiao2010smartdiet}. The system suggests personalized meals and restaurants based on the user’s daily nutritional requirements. To begin with, the system receives geographical coordinates from the user’s portable gadget and arranges with food suppliers adjoining to the user’s current area for securing accessible suppers and the corresponding nutrition actualities.

The majority of food journaling methods have a slow start and are difficult to maintain\cite{yang2017yum}. Therefore a great work has been conducted by Yang et al., 2017 who has developed a personalized nutrient-based meal recommendation system that tackles the limitation of the cold-start problem by not depending on the user’s dietary history \cite{yang2017yum}. They employ a food image analysis model called FoodDist in this system to learn about a user's fine-grained food preferences by interacting with them through food photographs and answering visually based diet-related queries. This allows them to create customised meal recommendations for anyone with dietary constraints such as vegetarian, vegan, kosher, or halal.

\section{Proposed AMRP System}

\begin{figure}[!h]
    \centering
    \includegraphics[scale=0.4]{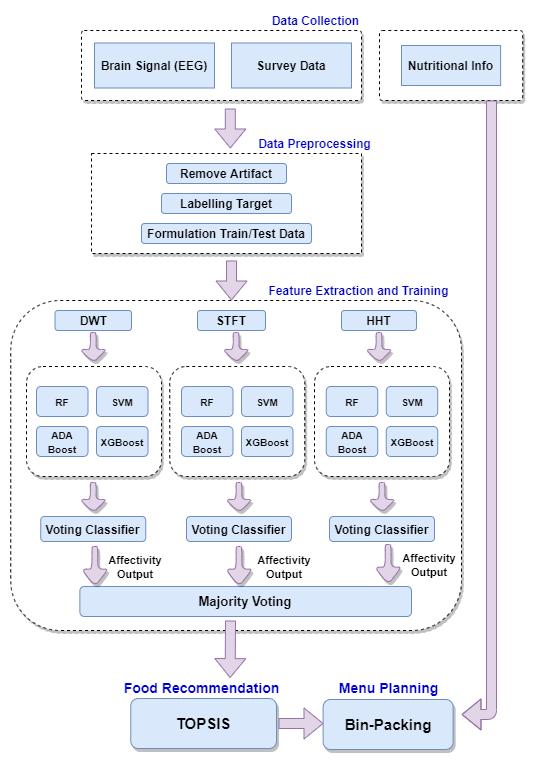}
    \caption{Modules of the Affective Behavior-based Personalized Meal Recommendation and Menu Planning (AMRP) Social System}
    \label{flowchart}
\end{figure}

The overview of the affective behavior-based personal-
ized meal recommendation and menu planning (AMRP)
system is presented in Figure~\ref{architecture}. Brain signals (EEG) and survey data are collected from participants as part of the data gathering procedure.Data from the survey includes responses to questions about how participants felt while watching a food image. The computing system is focused on receiving the raw signals, preprocess them, and computing affectivity. Food recommendations will demonstrate preference-based top 5 recommended foods for a person. Menu planning system recommends food for breakfast, lunch, dinner, and snacks which is a full-day menu plan. The decision of Menu planning is dependent on the nutrition of different foods along with affectivity. A complete workflow is demonstrated in Figure~\ref{flowchart}.

\subsection{Affective Computing for Food Recommendation}
Affective computing is comprised of EEG signal collection, survey data collection, preprocessing and feature extraction from raw signal, and finally model training. In this research, three feature extraction methods: Short-time Fourier transform (STFT), Discrete Wavelet Transform (DWT), Hilbert-Huang Transform (HHT) have been used along with four classifiers: Random Forest Classifier (RF), Support Vector Machine (SVM), XGBoost, and AdaBoost.

\subsubsection{Data Collection}
\begin{figure}[!h]
    \centering
    \includegraphics[scale=0.41]{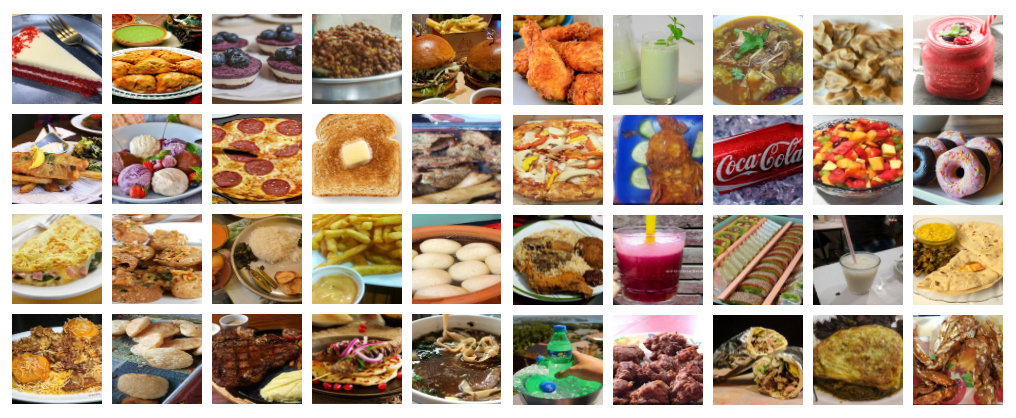}
    \caption{Pictures of foods used in the research}
    \label{foodImage}
\end{figure}

We collected EEG signals of 18 male and 7 females volunteers (25-40 years of age) with a 14 channel EMOTIV Epoc+ at a sampling rate of 128 HZ. A slideshow containing pictures of 40 different foods was used as stimuli (Figure~\ref{foodImage}). Each food picture was shown for 10 seconds. To overcome the effect of one picture to the next, we placed a calm scenario for 17 seconds after every food picture. In the mean time, the participants completed self assessment of Likeness, Valence and Arousal for the food.

We put the headset directly on the participant's scalp to record the EEG data. The whole environment was set up in a quiet and distraction-free area to obtain the participants' complete concentration, and they were also instructed to keep their muscle movements as minimum as possible.

A set of 40 images (Figure~\ref{foodImage}) were shown to the volunteers to record their neural response through EEG. These images were carefully chosen and set up in a sequential manner considering their nutrition value, popularity among certain age groups, and availability. The food slide covered heavy meals as well as light appetizers. All kinds of meals were presented, from common fast foods like burgers and pizza to everyday essentials like rice and chicken curry. A wide range of drinks choices like soft drinks, laban (watered down yoghurt drink), milk-saffron drink, etc., were also included in the slide to get an overall balanced response.

We asked our volunteers to answer a questionnaire after each food photo they saw. They had to rate their inclination surrounding that individual food, for \textbf{Likeness} ranging between 0 and 1 (0= Least, 1= Most). \textbf{Valance} as Excitement 0 and 1 (0= Least , 1= Most) and \textbf{Arousal} as Feelings labeled: \textit{Pleasant} as 1 and \textit{Disgust} as 0.

\subsubsection{Channel Selection}
EEG data was collected from two reference nodes (P3 and P4) and 14 channels (AF3, F7, F3, FC5, T7, P7, O1, O2, P8, T8, FC6, F4, F8). According to Coan et al.\cite{coan2001voluntary}, the frontal part of our brain is liable for positive and negative emotions. While the right frontal brain picks the negative emotions, the left frontal brain is associated with positive emotions. Therefore, throughout our study, we examined the signals of all 14 channels as well as the frontal channels (F3-F4, F7-F8, FC1-FC2, FC5-FC6) independently.
\subsubsection{Preprocessing}
Our target affectivity was categorized in three ways: Like (Least Like/Most Like), Excitement (Most Excitement/Least Excitement), and Feelings (Pleasant/Disgusted).


A [0.5-30] Hz band-pass channel was applied to the raw Signal \cite{zhang2021distilling}. In this research, we used the Wavelet-based denoising \cite{wavelet_new} method to segment the raw EEG data as in the Table \ref{frequency} into 5 groups with a window of 1 second. The wavelet-based denoising method is quite useful because it lets users precisely apply each wavelet-based sub-band threshold.

\begin{table}[!h]
\caption{EEG Signal Decomposition}
\label{frequency}
\centering
\begin{tabular}{|l || r|}
\hline
Bands & Frequency bands (Hz)\\
\hline
\hline
Delta & 0.3-4 \\
\hline
THETA & 4-8\\
\hline
ALPHA & 8-12\\
\hline
BETA & 12-25\\
\hline
GAMMA & 25-45\\
\hline
\end{tabular}

\end{table}
 
\subsubsection{Feature Extraction}
In a research conducted by (Al-Fahoum Al-Fraihat, 2014), it has been stated that signal noise which is captured during the recording of EEG signals, can affect the meaningful features of the signal \cite{al2014methods}. According to the studies of (Cvetkovic, Übeyli \& Cosic, 2008), features are used to recognize patterns without the loss of information being as low as possible \cite{cvetkovic2008wavelet}. Feature extraction is the method of recognizing patterns in a signal.
\begin{figure}[!h]
    \centering
    \includegraphics[scale=0.38]{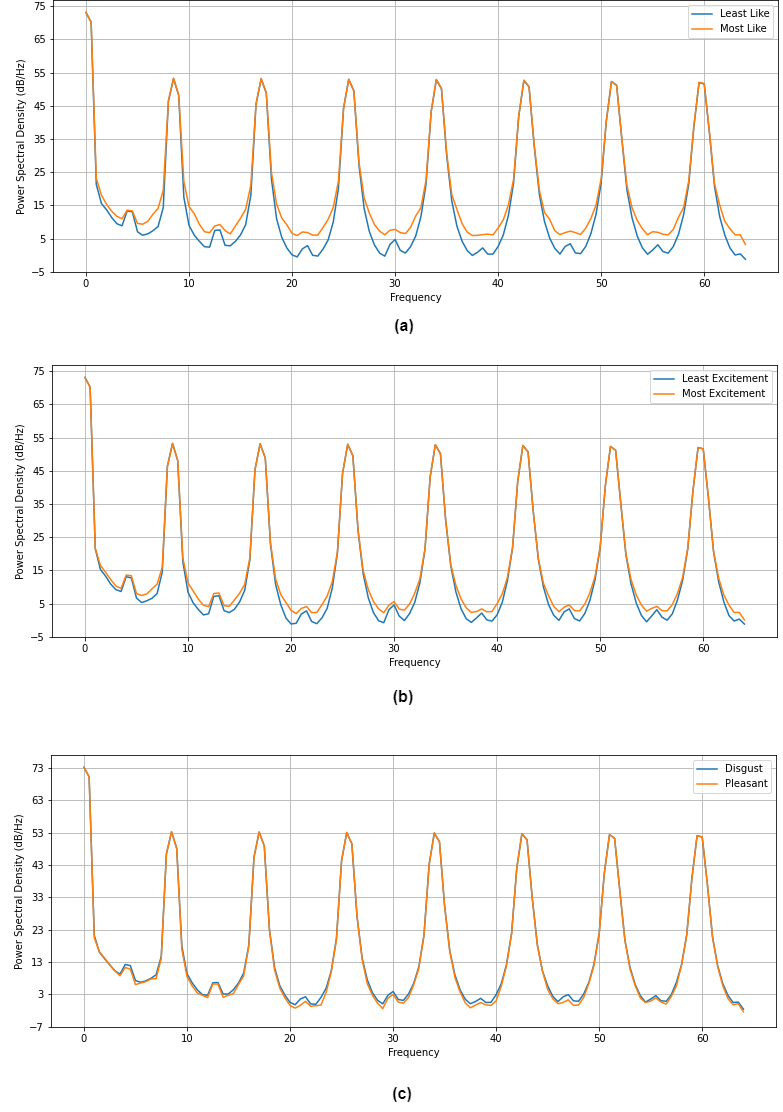}
    \caption{Power Spectral Density(PSD) for Like, Excitement, Feelings}
    \label{psdimg}
\end{figure}

In this research, we used three feature extraction methods: Discrete Wavelet Transform(DWT), Short Time Fourier Transform(STFT), and with Hilbert-Huang Transform(HHT). These extraction methods helped us to reduce the number of features presented in our obtained dataset.

When a signal is analyzed, it is converted into different frequencies like delta, alpha, beta, gamma, and theta. It represents the change of phase and magnitude in the frequency bands. The power spectral density (PSD) of the EEG signal is shown in Figure~\ref{psdimg}.

\begin{figure}[h]
    \centering
    \includegraphics[scale=0.4]{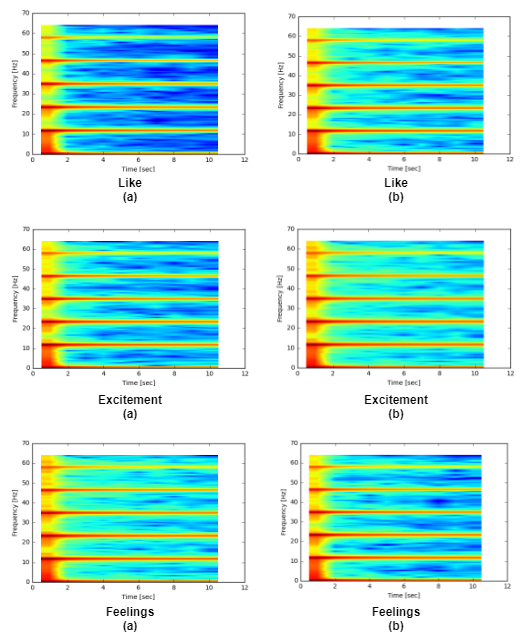}
    \caption{Spectrogram of STFT feature extraction for first 10 seconds}
    \label{stftimg}
\end{figure}

\paragraph{Short Time Fourier Transform (STFT)}
The sinusoidal frequency of a signal has been determined using this feature extraction method over time. This method is applied to specify the complexity of the amplitude of a signal against the time and frequency of a signal. The method entails performing a Fourier transform on a tiny segment of data at a time, which converts the signal into a 2D time and frequency function. This process can be expressed through the following equation \ref{eqbnstft}. Figure~\ref{stftimg} demonstrates spectrograms of the STFT feature extraction method for different ratings of three affectivity. For example, Like (a) is rated as 1, and Like (b) is rated as 2. Similarly, Excitement (a) is rated as 1, Excitement (b) is rated as 2. And Feelings (a) is rated as 0 and Feelings (b) is rated as 1.
\begin{equation}
\label{eqbnstft}
X(m, \omega)=\sum_{n=-\infty}^{\infty} x_{n} \omega_{n-m} e^{-i \omega t_{n}}
\end{equation}

\begin{figure}[h]
    \centering
    \includegraphics[scale=0.33]{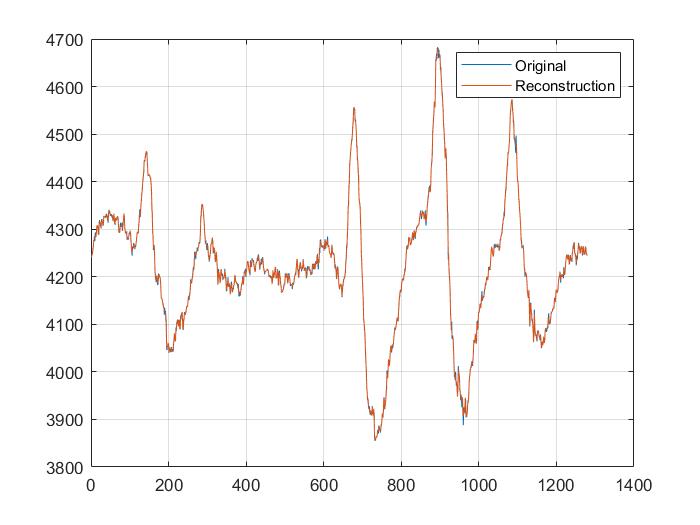}
    \caption{Raw Signal and reconstructed signal using DWT of first 10 seconds }
    \label{dwtimg}
\end{figure}

\paragraph{Discrete Wavelet Transform (DWT)}
Studies published by (Rankine, Mesbah Boashash, 2007) (Derya Ubeyli, 2008) states that DWT is the most favourable time-frequency resolution among all frequency ranges\cite{rankine2007matching}\cite{ubeyli2008analysis}. Long time windows are used in resolution to acquire a smaller low-frequency DWT, whereas short time windows are used in resolution to collect high-frequency data \cite{al2014methods}.
The DWT of a signal $x$ is calculated by passing it through a series of filters. The approximation coefficients are then split into a coarser approximation (low-pass) and high-pass (detail) component at each subsequent step.  The samples are first convolutioned by passing them through a low pass filter with an impulse response. This extraction method can be written as \ref{eqndwt}. Figure~\ref{dwtimg} shows the raw signal for the first 10 seconds and its corresponding wavelet signal. 
\begin{equation} \label{eqndwt} \psi(x)=\sum_{k=-\infty}^{\infty}(-1)^{k} a_{N-1-k} \psi(2 x-k)\end{equation}


\paragraph{Hilbert-Huang Transform (HHT)}
HHT is a two step method that uses Empirical Mode Decomposition (EMD) method to decompose the signal into Intrinsic Mode Functions (IMF) and applies Hilbert Spectral Analysis (HSA) to get the frequency data. Sifting is the process of extracting IMF from the EMD components. In our research, we used the HHT method to obtain the energy frequency-time distribution. This can be expressed as the following equation \ref{eqnhht}. Hilbert spectrum using HHT feature extraction from first 10 seconds of RAW signal is shown in Figure~\ref{hhtimg} 
\begin{equation} \label{eqnhht} x_{a}=F^{-1}(F(x) 2 U)=x+i y\end{equation}%

\begin{center}

\begin{minipage}{0.46\textwidth}
\begin{figure}[H]
    \centering
    \includegraphics[scale=0.25]{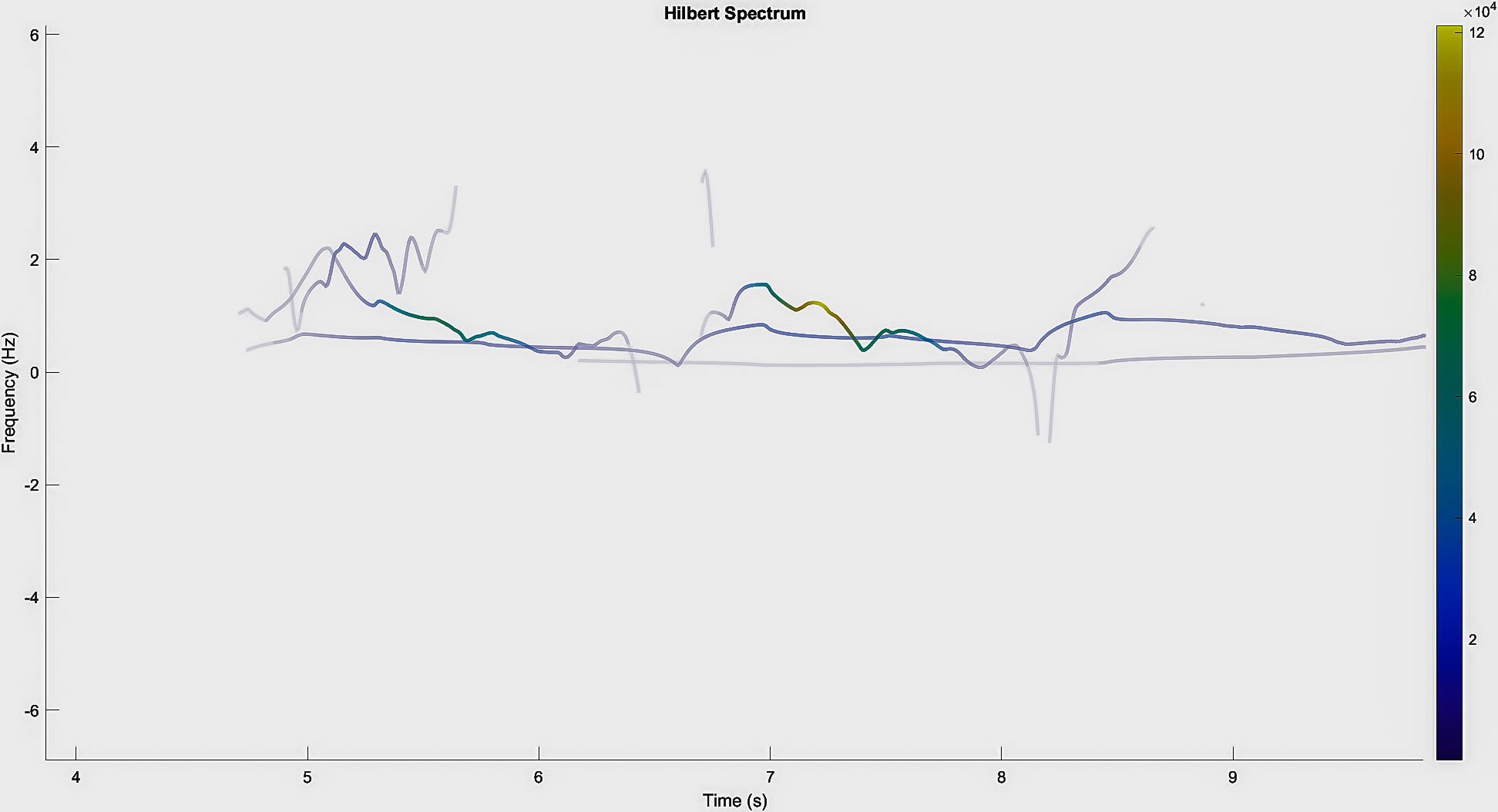}
    \caption{HHT spectrum for first 10 seconds of signal}
    \label{hhtimg}
\end{figure}
\end{minipage}

\end{center}


\subsubsection{Model Selection}
In this study, we used the Random Forest Classifier (RF), Support Vector Machine (SVM), XGBoost, and AdaBoost machine learning algorithms. Additionally, we used \textit{Scikit Learn Voting Classifier} to provide output that most classifiers could classify. In this study, we compared the results of affectivity using all 14 channels and only frontal channels which are F3-F4, F7-F8, FC1-FC2, FC5- FC6. Table \ref{allclassifiers} demonstrates comparison of accuracy and AUC score of different classifiers. Table \ref{confLike}, \ref{confExcitement}, \ref{confFeelings}, \ref{confLikefrontal}, \ref{confExcitementfrontal}, \ref{confFeelingsfrontal} shows confusion matrix and F1 score when we used voting classifier to get the output of the most occurrence from the classifiers by using all feature extraction methods separately.





\begin{table*}

\caption{Accuracy and AUC score by all classifiers using all channels (left table) and frontal channels (right table)}
\begin{minipage}[b]{0.45\linewidth}\centering

\begin{tabular}{|c|p{1cm}|p{1.6cm}|c|c|}

\hline Affectivity & Feature Extraction & Model & Accuracy & AUC \\
\hline 

\multirow{17}{*}{Like} & \multirow{5}{*}{DWT} & Random Forest &  90.29\% & 0.9031\\

\cline { 3 - 5 } &  & SVM  & 65.43\%&0.6555 \\
\cline { 3 - 5 } &  & XGBoost  &88.73\% & 0.8871 \\
\cline { 3 - 5 } &  & AdaBoost  & 75.14\%& 0.7506 \\
\cline { 3 - 5 } &  & Majority Voting  & 86.23\% & 0.8605 \\

\cline { 2 - 5 } & \multirow{5}{*}{STFT} & Random Forest & 48.93\%   & 0.4891 \\
 \cline { 3 - 5 } &  & SVM  & 48.93\% & 0.4891\\
\cline { 3 - 5 } &  & XGBoost  & 49.32\%& 0.4931\\
\cline { 3 - 5 } &  & AdaBoost  &52.81\% & 0.5286\\
\cline { 3 - 5 } &  & Majority Voting  &48.06\% & 0.4801\\

\cline { 2 - 5 } & \multirow{5}{*}{HHT} & Random Forest & 80.38\%  & 0.8031 \\
 \cline { 3 - 5 } &  & SVM  &57.28\% &0.5757 \\
\cline { 3 - 5 } &  & XGBoost  & 86.01\%& 0.8601\\
\cline { 3 - 5 } &  & AdaBoost  & 68.54\% & 0.6854 \\
\cline { 3 - 5 } &  & Majority Voting  & 83.82\%&0.8347 \\
\hline
\hline
\multirow{17}{*}{Excitement} & \multirow{5}{*}{DWT} & Random Forest & 76.05\%  &0.7627\\
\cline { 3 - 5 } &  & SVM  & 52.77\% & 0.5453 \\
\cline { 3 - 5 } &  & XGBoost  & 75.83\%&0.7591 \\
\cline { 3 - 5 } &  & AdaBoost  & 68.29\%&0.6855 \\
\cline { 3 - 5 } &  & Majority Voting  & 70.12\%&0.7029 \\

\cline { 2 - 5 } & \multirow{5}{*}{STFT} & Random Forest & 51.88\%  & 0.5204 \\
 \cline { 3 - 5 } &  & SVM  & 52.99\% &0.5413 \\
\cline { 3 - 5 } &  & XGBoost  & 54.10\%&0.5439 \\
\cline { 3 - 5 } &  & AdaBoost  & 52.77\% &0.5320 \\
\cline { 3 - 5 } &  & Majority Voting  & 52.72\%&0.5289 \\

\cline { 2 - 5 } & \multirow{5}{*}{HHT} & Random Forest & 72.94\%  & 0.7321 \\
 \cline { 3 - 5 } &  & SVM  &55.21\% & 0.5536 \\
\cline { 3 - 5 } &  & XGBoost & 76.71\%&0.7680 \\
\cline { 3 - 5 } &  &AdaBoost  & 67.40\%&0.6771 \\
\cline { 3 - 5 } &  & Majority Voting  &73.50\% &0.7365 \\
\hline

\hline
\hline
\multirow{17}{*}{Feelings} & \multirow{5}{*}{DWT} & Random Forest & 89.71\%  & 0.8972\\
\cline { 3 - 5 } &  & SVM  & 63.92\%&0.6411 \\
\cline { 3 - 5 } &  & XGBoost& 87.47\%&0.8758 \\
\cline { 3 - 5 } &  & AdaBoost  & 74.95\%&0.7531 \\
\cline { 3 - 5 } &  & Majority Voting  &84.82\% &0.8479 \\

\cline { 2 - 5 } & \multirow{5}{*}{STFT} & Random Forest &  63.73\% &0.6365 \\
 \cline { 3 - 5 } &  & SVM  &49.53\% &0.4987 \\
\cline { 3 - 5 } &  & XGBoost& 65.79\% &0.6576 \\
\cline { 3 - 5 } &  & AdaBoost  & 51.77\%&0.5182 \\
\cline { 3 - 5 } &  & Majority Voting  &43.75\% & 0.4368\\

\cline { 2 - 5 } & \multirow{5}{*}{HHT} & Random Forest &  78.87\% & 0.7927 \\
 \cline { 3 - 5 } &  & SVM  &62.24\% & 0.6251 \\
\cline { 3 - 5 } &  & XGBoost& 85.79\%& 0.8597\\
\cline { 3 - 5 } &  & AdaBoost  & 71.96\%&0.7221 \\
\cline { 3 - 5 } &  & Majority Voting  & 79.24\%&0.7913 \\
\hline
\end{tabular}
\end{minipage}
\hspace{0.5cm}
\begin{minipage}[b]{0.45\linewidth}
\centering

\begin{tabular}{|c|p{1cm}|p{1.6cm}|c|c|}

\hline Affectivity & Feature Extraction & Model & Accuracy & AUC \\
\hline 

\multirow{17}{*}{Like} & \multirow{5}{*}{DWT} & Random Forest & 86.79\%  & 0.8685\\

\cline { 3 - 5 } &  & SVM  & 59.61\%& 0.5915\\
\cline { 3 - 5 } &  &  XGBoost &83.68\% & 0.8371 \\
\cline { 3 - 5 } &  & AdaBoost  & 70.48\%&0.7042 \\
\cline { 3 - 5 } &  & Majority Voting  & 78.01\% & 0.7796 \\

\cline { 2 - 5 } & \multirow{5}{*}{STFT} & Random Forest &  47.18\% & 0.4713 \\
 \cline { 3 - 5 } &  & SVM  & 46.40\% & 0.46.92 \\
\cline { 3 - 5 } &  & XGBoost  & 53.20\% & 0.5323 \\
\cline { 3 - 5 } &  & AdaBoost  & 48.15\% & 0.4820 \\
\cline { 3 - 5 } &  & Majority Voting  &50.77\%&0.5069 \\

\cline { 2 - 5 } & \multirow{5}{*}{HHT} & Random Forest & 86.79\%  & 0.8684 \\
 \cline { 3 - 5 } &  & SVM  & 55.72\% & 0.5522 \\
\cline { 3 - 5 } &  & XGBoost  & 83.88\%& 0.8389\\
\cline { 3 - 5 } &  & AdaBoost  & 64.07\% & 0.6398 \\
\cline { 3 - 5 } &  & Majority Voting  &77.43\% &0.7738 \\
\hline
\hline
\multirow{17}{*}{Excitement} & \multirow{5}{*}{DWT} & Random Forest &  75.60\% & 0.7550\\
\cline { 3 - 5 } &  & SVM  & 53.88\% & 0.5566 \\
\cline { 3 - 5 } &  & XGBoost  & 76.49\%& 0.7644\\
\cline { 3 - 5 } &  & AdaBoost  & 65.63\%&0.6593 \\
\cline { 3 - 5 } &  & Majority Voting  &70.31\% &0.7055 \\

\cline { 2 - 5 } & \multirow{5}{*}{STFT} & Random Forest & 55.43\%  & 0.5572 \\
 \cline { 3 - 5 } &  & SVM  & 49.66\% &0.5196 \\
\cline { 3 - 5 } &  & XGBoost  & 55.21\%&0.5531 \\
\cline { 3 - 5 } &  & AdaBoost  & 54.98\% &0.5501 \\
\cline { 3 - 5 } &  & Majority Voting  & 46.65\%&0.4704 \\

\cline { 2 - 5 } & \multirow{5}{*}{HHT} & Random Forest &  70.73\% & 0.7064\\
 \cline { 3 - 5 } &  & SVM  & 50.99\% & 0.5283 \\
\cline { 3 - 5 } &  & XGBoost  & 73.83\% & 0.7382 \\
\cline { 3 - 5 } &  & AdaBoost  & 62.30\%& 0.6275\\
\cline { 3 - 5 } &  & Majority Voting  & 67.41\% &0.6772 \\
\hline

\hline
\hline
\multirow{17}{*}{Feelings} & \multirow{5}{*}{DWT} & Random Forest & 90.28\%  & 0.9031\\
\cline { 3 - 5 } &  & SVM  & 58.13\%&0.5959 \\
\cline { 3 - 5 } &  & XGBoost  & 87.47\%&0.8766 \\
\cline { 3 - 5 } &  & AdaBoost  & 72.14\% & 0.7258 \\
\cline { 3 - 5 } &  & Majority Voting  &82.73\% &0.8294 \\

\cline { 2 - 5 } & \multirow{5}{*}{STFT} & Random Forest &  56.07\% & 0.5654 \\
 \cline { 3 - 5 } &  & SVM  & 50.84\%&0.5273 \\
\cline { 3 - 5 } &  & XGBoost  & 55.32\%&0.5564 \\
\cline { 3 - 5 } &  & AdaBoost  & 51.58\% & 0.5184 \\
\cline { 3 - 5 } &  & Majority Voting  &52.15\% &0.5262 \\

\cline { 2 - 5 } & \multirow{5}{*}{HHT} & Random Forest & 88.59\%  &0.8852 \\
 \cline { 3 - 5 } &  & SVM  & 54.95\%& 0.5661\\
\cline { 3 - 5 } &  & XGBoost  & 87.10\%&0.8725 \\
\cline { 3 - 5 } &  & AdaBoost  & 71.21\%&0.7148 \\
\cline { 3 - 5 } &  & Majority Voting  & 84.80\%&0.8489 \\
\hline
\end{tabular}
\end{minipage}
\label{allclassifiers}
\end{table*}

\begin{table}[H]
\caption{Confusion Matrix \& F1 Score for Like Affectivity by Voting Classifier for all channels.}
\centering
\begin{tabular}{|c||c|c|c|c|}

\hline 
 & Actual Class & \multicolumn{2}{c|} { Predicted Class } & F1 Score \\
   \cline {3-4 } & & Least Like & Most Like & \\
\hline
\hline 

\multirow{2}{*}{DWT}&Least Like & 214 & 13 & 0.877 \\
&Most Like & 47 & 165 & \\
\hline
\hline
\multirow{2}{*}{STFT}&Least Like & 112   & 115 & 0.4955 \\
&Most Like & 113   & 99 & \\
\hline
\hline
\multirow{2}{*}{HHT}& Least Like & 213    & 14 & 0.8571 \\
&Most Like & 57   & 155 & \\
\hline

\end{tabular}
\label{confLike}

\end{table}

From TABLE \ref{confLike}, we observe that F1 score is the highest when using DWT feature extraction to classify like affectivity.

\begin{table}[H]
\caption{Confusion Matrix \& F1 Score for Excitement Affectivity by Voting Classifier for all channels. [ L.E = Least Excitement, M.E = Most Excitement ]}
\centering
\begin{tabular}{|c||c|c|c|c|}

\hline 
 & Actual Class & \multicolumn{2}{c|} { Predicted Class } & F1 Score \\
   \cline {3-4 } & &  L.E &  M.E & \\
\hline
\hline 

\multirow{2}{*}{DWT}& L.E & 150 & 39 & 0.7228 \\
& M.E & 76 & 120 & \\
\hline
\hline
\multirow{2}{*}{STFT}& L.E & 117   & 72 & 0.5625 \\
&  M.E & 110   & 86 & \\
\hline
\hline
\multirow{2}{*}{HHT}&  L.E & 155    & 34 & 0.7524 \\
& M.E & 68   & 128 & \\
\hline

\end{tabular}
\label{confExcitement}
\end{table}

From TABLE \ref{confExcitement}, when classifying excitement affectivity, HHT performs well amongst other feature extraction method. TABLE \ref{confFeelings} shows that higher F1 score is achieved by the voting classifier while considering DWT feature extraction method for classifying Feelings affectivity.

\begin{table}[H]
\caption{Confusion Matrix \& F1 Score for Feelings Affectivity by Voting Classifier for all channels. }
\centering
\begin{tabular}{|c||c|c|c|c|}

\hline 
 & Actual Class & \multicolumn{2}{c|} { Predicted Class } & F1 Score \\
   \cline {3-4 } & & Disgust & Pleasant & \\
\hline
\hline 

\multirow{2}{*}{DWT}&Disgust & 198 & 28 & 0.8425 \\
&Pleasant & 40 & 182 & \\
\hline
\hline
\multirow{2}{*}{STFT}&Disgust & 115   & 111 & 0.3913 \\
&Pleasant & 141   & 81 & \\
\hline
\hline
\multirow{2}{*}{HHT}& Disgust & 206    & 20 & 0.7621 \\
&Pleasant & 73   & 149 & \\
\hline

\end{tabular}
\label{confFeelings}

\end{table}

\begin{table}[H]
\caption{Confusion Matrix \& F1 Score using Voting Classifier for Like Affectivity for only frontal channels.}
\centering
\begin{tabular}{|c||c|c|c|c|}

\hline 
 & Actual Class & \multicolumn{2}{c|} { Predicted Class } & F1 Score \\
   \cline {3-4 } & & Least Like & Most Like & \\
\hline
\hline 

\multirow{2}{*}{DWT}&Least Like & 236  & 22 & 0.8068 \\
&Most Like & 91  & 165 & \\
\hline
\hline
\multirow{2}{*}{STFT}&Least Like & 188   & 70 & 0.5977 \\
&Most Like & 183    & 73 & \\
\hline
\hline
\multirow{2}{*}{HHT}& Least Like & 233  & 25 & 0.80 \\
&Most Like & 91   & 165 & \\
\hline

\end{tabular}
\label{confLikefrontal}

\end{table}

Table \ref{confLikefrontal}, \ref{confExcitementfrontal}, and \ref{confFeelingsfrontal} shows confusion matrix and F1 score for Like affectivity, Excitement affectivity, and Feelings affectivity respectively when we considered only frontal channels in our experiment. When the DWT feature extraction approach was taken into consideration, Like affectivity and Excitement affectivity had the greatest F1 score, while Feelings affectivity received the best F1 score when the HHT feature extraction method was taken into account.

\begin{table}[H]
\caption{Confusion Matrix \& F1 Score for Excitement Affectivity by Voting Classifier for only frontal channels. [ L.E = Least Excitement, M.E = Most Excitement ]}
\centering
\begin{tabular}{|c||c|c|c|c|}

\hline 
 & Actual Class & \multicolumn{2}{c|} { Predicted Class } & F1 Score \\
   \cline {3-4 } & & L.E & M.E & \\
\hline
\hline 

\multirow{2}{*}{DWT}& L.E & 152  & 68 & 0.7355 \\
&M.E & 171  & 57 & \\
\hline
\hline
\multirow{2}{*}{STFT}& L.E & 152   & 68 & 0.5598 \\
&M.E & 171     & 57 & \\
\hline
\hline
\multirow{2}{*}{HHT}&  L.E & 188  & 32 & 0.7203\\
&M.E & 114     & 114 & \\
\hline

\end{tabular}
\label{confExcitementfrontal}

\end{table}

\begin{table}[H]
\caption{Confusion Matrix \& F1 Score for Feelings Affectivity by Voting Classifier for only frontal channels.}
\centering
\begin{tabular}{|c||c|c|c|c|}

\hline 
 & Actual Class & \multicolumn{2}{c|} { Predicted Class } & F1 Score \\
   \cline {3-4 } & & Disgust & Pleasant & \\
\hline
\hline 

\multirow{2}{*}{DWT}&Disgust & 250   & 12 & 0.8059 \\
&Pleasant & 80  & 191 & \\
\hline
\hline
\multirow{2}{*}{STFT}&Disgust & 211     & 51 & 0.3444\\
&Pleasant & 204    & 67 & \\
\hline
\hline
\multirow{2}{*}{HHT}& Disgust & 236      & 26 & 0.8421  \\
&Pleasant & 55    & 216 & \\
\hline

\end{tabular}
\label{confFeelingsfrontal}

\end{table}

 While comparing confusion matrix between all channels and only frontal channels, TABLE~\ref{confLike}, \ref{confExcitement}, \ref{confFeelings}, \ref{confLikefrontal}, \ref{confExcitementfrontal}, \ref{confFeelingsfrontal} shows that classification using all channels achieves better F1-Score than using only frontal channels.

\paragraph{Hierarchical Ensembling}
In our proposed work, we have used \textit{Scikit-Learn Voting Classifier} on top of four classifiers: RF, SVM, AdaBoost, and XGBoost to get mostly occured targets classified by these algorithms using  three feature extraction methods separately. Then we combined the outputs and again applied the ensemble technique to get the final classification. In order to decrease the likelihood of overfitting, hierarchical ensembling was incorporated to classify affectivity.

\subsection{Food Recommendation Using TOPSIS}

Yoon and Hwang introduced the TOPSIS approach for tackling Multiple Criteria Decision Making in 1980. (MCDM). The Euclidean distance between the alternative and ideal solution is calculated to determine the best alternative option. Positive Ideal Solution (PIS) uses the smallest distance, whereas Negative Ideal Solution (NIS) uses the longest distance (NIS).

Step 1: With $n$ index of evaluation and $m$ evaluation for indexes, its decision matrix A is:

\newcommand{\indsize}{\scriptsize}
\newcommand{\colind}[2]{\displaystyle\smash{\mathop{#1}^{\raisebox{.5\normalbaselineskip}{\indsize #2}}}}
\newcommand{\rowind}[1]{\mbox{\indsize #1}}

$$
  A =
  \begin{array}{@{}c@{}}
    \rowind{F\textsubscript{1}} \\ \rowind{F\textsubscript{2}} \\ \rowind{\vdots} \\ \rowind{F\textsubscript{m}} 
  \end{array}
  \mathop{\left[
  \begin{array}{ *{5}{c} }
     \colind{x\textsubscript{11}}{C\textsubscript{1}}  &  \colind{\dots}{\dots}  &  \colind{x\textsubscript{13}}{C\textsubscript{n}} \\
     
     x\textsubscript{21} &  \dots  & x\textsubscript{23}   \\
     \vdots & \vdots & \vdots\\
      x\textsubscript{n1} &  \dots  & x\textsubscript{n3} 
     
  \end{array}
  \right]}^{
  }
$$

Where $C$ denotes criteria, index $C=1...n$ denotes the number of criteria.This research employs three criteria: Like Affectivity, Excitement Affectivity and Feelings Affectivity. $F$ denotes an alternate solution, and index $F=1...m$ denotes the number of alternative solutions.In our research, $F_m=40$ because there are 40 items.

Normalize the decision matrix A using this formula :

\begin{equation}
\mathrm{NDM}=\mathrm{R}_{\mathrm{ij}}=\frac{\mathrm{x}_{\mathrm{ij}}}{\sqrt{\sum_{\mathrm{i}=1}^{\mathrm{m}} \mathrm{x}_{\mathrm{ij}}^2}}
\end{equation}
  
Step 2: Constructing a Weighted Normalized Matrix $V$ with weight vector,\\ $$W ={w\textsubscript{1},w\textsubscript{2}, ..... , w\textsubscript{n},}$$
\begin{equation}
\begin{gathered}
V=\left[\begin{array}{cccc}
v_{11} & v_{12} & \ldots & v_{1 m} \\
v_{21} & v_{22} & \ldots & v_{2 m} \\
\ldots & \ldots & \ldots & \ldots \\
v_{n 1} & v_{n 2} & \ldots & v_{n m}
\end{array}\right] 
\end{gathered}
\end{equation}
\\
$$v_{i j}=w_{j} ; R_{i j}, i=1,2, \ldots, n ; j=1,2, \ldots m$$

\begin{equation}
\mathrm{V}=\mathrm{V}_{\mathrm{ij}}=\mathrm{W}_{\mathrm{j}} \times \mathrm{R}_{\mathrm{ij}-\mathrm{u}}
\end{equation}
  
Step 3: Calculate the Positive and Negative Ideal Solution Values.

Step 4: Determine how far each competing alternative is from the ideal and non-ideal solution. Where, $i$ = criterion index, $j$ = alternative index.

\begin{equation}
\begin{aligned}
&\mathrm{S}^{+}=\sqrt{\sum_{j=1}^{n}\left(V_{j}^{+}-V_{i j}\right)^{2} \mathrm{i}}=1 \ldots \ldots, \mathrm{m} \\
&\mathrm{S}^{-}=\sqrt{\sum_{j=1}^{n}\left(V_{j}^{-}-V_{i j}\right)^{2} \mathrm{i}}=1 \ldots \ldots, \mathrm{m}
\end{aligned}
\end{equation}
 
Step 5: Determine the distance between each site and the best solution. Relative proximity of the possible site to the optimum solution is determined for each competitive alternative.

\begin{equation}
\mathrm{Ci}=\mathrm{S}_{\mathrm{i}} /\left(\mathrm{S}_{\mathrm{i}}^{+}+\mathrm{S}_{\mathrm{i}}\right), 0 \leq \mathrm{Ci} \leq 1
\end{equation}
 
Step 6(Final Step): Using the value of C, arrange all potential alternative solutions. First rank for the more excellent value, which has a solution based on relative proximity.

\subsection{Menu Planing Using Bin-Packing Algorithm}
In the bin-packing issue, we are given a list $L$ of real values ranging from 0 to 1 that we must distribute to unit capacity bins in such a way that no bin receives more than 1 and a minimal number of bins is needed.  It is a combinatorial NP-hard problem according to computational complexity theory \cite{trattner2017investigating}. The choice problem (determining whether or not things will fit into a given number of bins) is $NP-complete$ \cite{harvey2012learning}.
If there are $n$ items and $n$ bins, with 
\begin{equation}
w_{j} = \text{weight of item } j
\end{equation}
\begin{equation}
c = \text{capacity of each bin} 
\end{equation}

assign each item to its own container such that the total weight of the items in each bin does not exceed $c$ and the number of bins used is kept to a minimum. The following is a mathematical version of the problem that has been offered.

$$
\begin{aligned}
\operatorname{minimize, } & z=\sum_{i=1}^{k} y_{i} \\
\text { subject to } & \sum_{j=1}^{n} w_{j} x_{i j} \leq c y_{i}, \quad i \in N=\{1, \ldots, n\}, \\
& \sum_{i=1}^{n} x_{i j}=1, \quad j \in N, \\
& y_{i}=0 \text { or } 1, \quad i \in N, \\
& x_{i j}=0 \text { or } 1, \quad i \in N, j \in N,\\
\text{where}\\
y_{i} &=\left\{\begin{array}{ll}
1 & \text { bin } i \text { is used; } \\
0 & \text { otherwise }
\end{array}\right.\\
x_{i j} &=\left\{\begin{array}{ll}
1 & \text { item } j \text { is assigned to bin } i ; \\
0 & \text { otherwise }
\end{array}\right.
\end{aligned}
$$

\begin{center}

\begin{minipage}{0.46\textwidth}
\begin{algorithm}[H]
\caption{\textbf{:} Algorithm for Affective Computing Based
Food Recommendation and Menu Planning}
\label{alg1}
\begin{algorithmic}[1]
\renewcommand{\algorithmicrequire}{\textbf{Input:}}
\renewcommand{\algorithmicensure}{\textbf{Output:}}
\REQUIRE CSV Data from EEG Signal, Nutritional data
\ENSURE Affectivity, Food Recommendation, Menu Plan
\STATE Remove artifacts from raw csv
\FOR{affective in Likeness, Excitement and Feelings}
\STATE Feature Extraction (STFT,DWT,HHT)
\FOR{STFT,DWT,HHT feature extraction}
\STATE Voting Classifier (RF, SVM, AdaBoost, XGBoost)
\ENDFOR
\STATE $Output=$Majority Voting $(STFT, DWT, HHT)$
\STATE AffectiveResult.append($Output$)
\ENDFOR
\STATE Recommended 5 Food = TOPSIS (Likeness, Exceitement, Feelings)
\STATE Menu Plan = Bin Packing (Breakfast, Lunch, Snacks, Dinner, Calorie)
\end{algorithmic}
\end{algorithm}
\end{minipage}
\end{center}








In this research, we have planned a menu planning system for breakfast, lunch, dinner, and snacks. While using the Bin-packing algorithm, nutritional information is the key to plan a menu. We have constructed a bin for each meal, and the algorithm generates a whole menu plan based on the suggested calorie consumption. This paper proposes Algorithm~\ref{alg1} which is the overall insight of this research. After gathering the required data, affectivity was classified by majority voting of the predicted output of four classifiers using three different feature extraction techniques following hierarchical ensembling methodology, and the TOPSIS method was then used on top of the affectivity to choose the meals to propose. To arrange the whole day menu, Bin-packing algorithm is employed.

\section{Performance Evaluation}
\subsection{Implementation}
All the models are trained with 70\% of the data and the rest of the data is for testing purpose. In this research, we have proposed subject independent analysis.
\subsection{Evaluation Metrics}
Evaluation metric determines the performance of a predictive model. In this paper, 3 evaluation metrics: accuracy, confusion matrix, f1-Score, AUC score are used to evaluate our models.
\subsubsection{Confusion Matrix}
A confusion matrix is a table that is frequently used to specify the output of a classification model or classifier on a set of test data that has known true values. The output of the matrix is made up of four values, True Positive (TP), True Negative (TN), False Positive (FP), and False Negative (FN).
\\
\textbf{True Positive:}When the value of test information for which actual values are known is shown as true positive, the value is shown as true positive.
\\
\textbf{False Positive:} False positive occurs when the value of the classifier is showing positive results while the actual or true class value is negative.
\\
\textbf{True Negative:} It will be a true negative if the predicted class is negative and the actual value is also negative.
\\
\textbf{False Negative:} The output will be false negative if the actual class is a positive result but the anticipated class is a negative outcome.
\\
Accuracy can be measured using following equation rate of our classifiers, that is how often the classifier has able to predict correctly and how often it shows wrong prediction. Considering \textit{n} as the total number of inputs given to the classifier, we can measure how often the classifier was right, by using the following equation:
\begin{equation}
\label{accuracy}
\text { Accuracy }=\frac{T P+T N}{n}
\end{equation}
And to measure the error rate or how often the classifier has been wrong in its prediction, the following equation was used:
\begin{equation}
   \text {Misclassification rate} =\frac{FP+FN}{n} 
\end{equation}

\subsubsection{F1 Score}
The F1-score is a method of combining the model's accuracy and recall, and it is defined as the harmonic mean of the model's precision and recall. The formula for the F1-score norm is the harmonic mean of precision and recall. A great model with an F-score of 1.
\begin{equation}
\mathrm{F} 1-\text { Score }=2 * \frac{\text {precision} \text { *true positive }}{\text {precision}+\text {true positive}}
\end{equation}
\subsubsection{AUC Score}
The AUC score evaluates how well a classifier can distinguish between classes. The score works as a measurement of classification at various points. Closely related to AUC is the ROC(Receiver Operator Characteristic) curve which is a probability curve. The TPR (True Positive Rate) is shown against the FPR (False Positive Rate) value produced from the confusion matrix in the ROC curve. The AUC score is derived from the ROC curve by measuring the area under it. The higher the AUC score of a model is, the better it performs at separating and classifying data. The best model has an AUC score of 1 which translates into it being a better classifier. A poor model has a score of 0 which means it misclassified the data completely. 

\subsection{Result Analysis}
In this section, we have demostrated the  results of affectve computing by hierarchical ensembling, and TOPSIS based food recommendation and menu planning.

\subsubsection{Affectivity Computing for Food Recommendation}


\begin{figure}[!ht]
    \centering
    \includegraphics[scale=0.35]{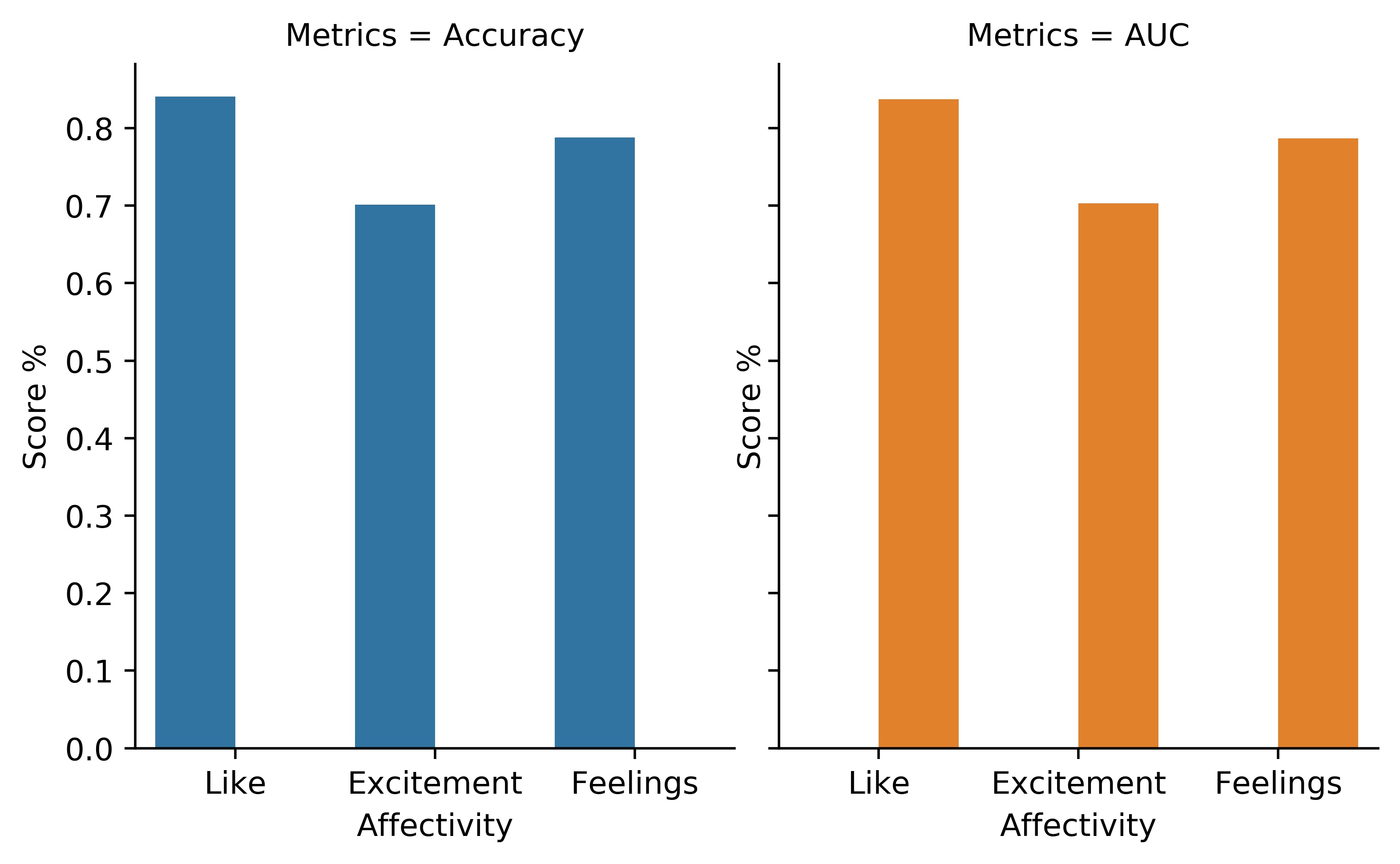}
    \caption{Accuracy and AUC after Hierarchical Ensembling using all channels}
    \label{majority_voting_all_channel}
\end{figure}
\begin{figure}[H]
    \centering
    \includegraphics[scale=0.35]{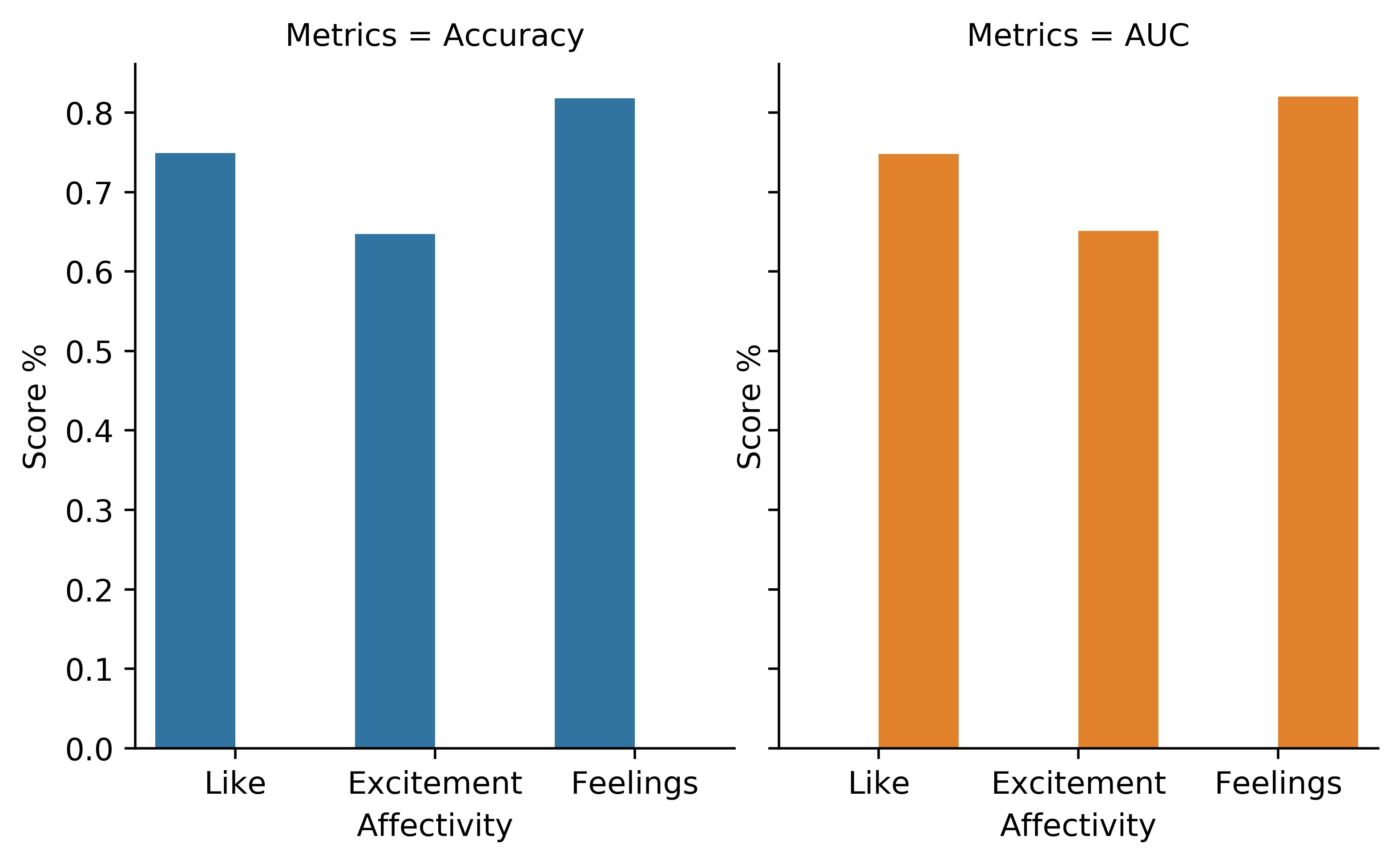}
    \caption{Accuracy and AUC after Hierarchical Ensembling using frontal channels [F3-F4, F7-F8, FC1-FC2, FC5- FC6]}
    \label{majority_voting_frontal_channel}
\end{figure}
In this research,  majority voting is used over results obtained from all feature extraction methods.
By analyzing Figure~\ref{majority_voting_all_channel}, \ref{majority_voting_frontal_channel} while considering all channels for \textit{Like Affectivity}, \textit{Excitement Affectivity}, \textit{Feelings Affectivity} both accuracy and AUC score were higher than trained model with frontal channels (\textit{F3-F4,  F7-F8,  FC1-FC2,  FC5- FC6}). In both cases, the differnce between  accuracy and AUC score is very low. Accuracy and AUC of \textit{Like Affectivity}, \textit{Excitement Affectivity} and \textit{Feelings Affectivity} were  \textbf{10.88\%}, \textbf{7.6\%}, \textbf{3.03\%} higher respectively while training model using all channels.


\begin{table}[H]
\caption{Confusion Matrix \& F1 Score For Like Affectivity by Hierarchical Ensembling}
\centering
\begin{tabular}{|p{2cm}||p{1.8cm}|p{1cm}|p{1cm}|p{0.7cm}|}

\hline 
 & Actual Class & \multicolumn{2}{c|} { Predicted Class } & F1 Score \\
   \cline {3-4 } & & Least Like & Most Like & \\
\hline
\hline 

\multirow{2}{*}{All Channels}&Least Like & 212 & 15 & 0.8582 \\
&Most Like & 55 & 157 & \\
\hline
\hline
\multirow{2}{*}{Frontal Channels}&Least Like & 237 & 21 & 0.7860 \\
&Most Like & 108   & 148 & \\
\hline
\end{tabular}
\label{confMajLike}
\end{table}

\begin{table}[H]
\caption{Confusion Matrix \& F1 Score For Excitement Affectivity by Hierarchical Ensembling [ L.E= Least Excitement, M.E= Most Excitement ]}
\centering
\begin{tabular}{|c||c|c|c|p{1.3cm}| }

\hline 
 & Actual Class & \multicolumn{2}{c|} { Predicted Class } & F1 Score \\
   \cline {3-4 } & & L.E & M.E & \\
\hline
\hline 

\multirow{2}{*}{All Channels}&L.E & 153 & 36 & 0.7268 \\
&M.E & 79 & 117 & \\
\hline
\hline
\multirow{2}{*}{Frontal Channels}&L.E & 192 & 28 & 0.7084 \\
&M.E & 130  & 98 & \\
\hline
\end{tabular}
\label{confMajExcitement}
\end{table}

\begin{table}[H]
\caption{Confusion Matrix \& F1 Score For Feelings Affectivity by Hierarchical Ensembling}
\centering
\begin{tabular}{|c||c|c|c|p{0.8cm}|}

\hline 
 & Actual Class & \multicolumn{2}{c|} { Predicted Class } & F1 Score \\
   \cline {3-4 } & & Disgust & Pleasant & \\
\hline
\hline 

\multirow{2}{*}{All Channels}&Disgust & 212 & 15 & 0.8582 \\
&Pleasant & 55 & 157 & \\
\hline
\hline
\multirow{2}{*}{Frontal Channels}&Disgust & 248 & 14 & 0.7949 \\
&Pleasant & 83   & 188 & \\
\hline
\end{tabular}
\label{confMajFeelings}
\end{table}
Observing Table~\ref{confMajLike}, \ref{confMajExcitement} and \ref{confMajFeelings}, when the model is trained with will all channels F1 Score is bit hugh for \textit{Like Affectivity} and \textit{Feelings Affectivity}. However, there is a significant difference for \textit{Excitement Affectivity} in case of F1 Score. 

To conclude, though frontal channels captures positive and negative emotion precisely, in this paper, based on various performance evaluation metrics, it has been observed that performance of using all channels outweighs using only frontal channels in terms of computing affectivity.

\subsubsection{Result Analysis of TOPSIS Based Food Recommendation}
Determinde decison matrix A :

$$
  A =
  \begin{array}{@{}c@{}}
    \rowind{F\textsubscript{1}}  \\ \rowind{F\textsubscript{2}} \\ \rowind{\vdots} \\ \rowind{F\textsubscript{40}} 
  \end{array}
  \mathop{\left[
  \begin{array}{ *{5}{c} }
     \colind{2}{Like}  &  \colind{2}{Excitement}  &  \colind{1}{Feelings} \\
     
     1 & 1 & 0\\
     \vdots & \vdots & \vdots\\
      2 &  2  & 1 
     
  \end{array}
  \right]}^{
  }
$$

Normalization Vector :
$$[ 11.6619, 11.401, 5.567 ]$$

Normalized Decision Matrix :

$$
  A =
  \begin{array}{@{}c@{}}
    \rowind{F\textsubscript{1}}  \\ \rowind{F\textsubscript{2}} \\ \rowind{\vdots} \\ \rowind{F\textsubscript{40}} 
  \end{array}
  \mathop{\left[
  \begin{array}{ *{5}{c} }
     \colind{0.1714}{Like}  &  \colind{0.1754}{Excitement}  &  \colind{0.1796}{Feelings} \\
     
     0.0857 & 0.0877 & 0\\
     \vdots & \vdots & \vdots\\
      0.1714 &  0.1754 & 0.1796
     
  \end{array}
  \right]}^{
  }
$$

Weight vector :
$$[ 0.4.0.3.0.3 ]$$

Weighted Normalized Decision Matrix:

$$
  A =
  \begin{array}{@{}c@{}}
    \rowind{F\textsubscript{1}}  \\ \rowind{F\textsubscript{2}} \\ \rowind{\vdots} \\ \rowind{F\textsubscript{40}} 
  \end{array}
  \mathop{\left[
  \begin{array}{ *{5}{c} }
     \colind{0.06859}{Like}  &  \colind{0.05262}{Excitement}  &  \colind{0.05388}{Feelings} \\
     
     0.03430 & 0.02631 & 0\\
     \vdots & \vdots & \vdots\\
      0.06859 &  0.05262 & 0.05388
     
  \end{array}
  \right]}^{
  }
  $$

Ideal best value :

$$\mathrm{V}_{\mathrm{j}}^{+} = [0.06859, 0.05262, 0.05388]$$

Ideal worst value :

$$\mathrm{V}_{\mathrm{j}}^{-} = [0.03429, 0.02631, 0]$$

Diatance from Ideal Best :
$$\mathrm{S}^{+} = 0, 0.0690, \dots ,0$$

Diatance from Ideal Worst :
$$\mathrm{S}^{-} = 0.0690, 0, \dots ,0.0690$$

\begin{table}[H]
\caption{Top 5 food choices for Person 1 (Male) }
\centering
\begin{tabular}{|c|c|c|}
\hline Serial No. & Food Name & Topsis Score \\
\hline 1 & Ramen & 0.706 \\
\hline 2 & Burger & 0.698 \\
\hline 3 & Beef Steak & 0.687 \\
\hline 4 & Kacchi Biriyani & 0.687 \\
\hline 5 & Polao-Roast & 0.666 \\
\hline
\end{tabular}
\label{fr1}
\end{table}

In this paper, we have implemented 3 affectivity criteria: Like, Feelings, and Excitement and there are 40 items of foods. These foods will be recommended to individual persons. From Table \ref{fr1},\ref{fr2} we can see that for different persons, the recommendation system shows  results sorted by \textit{Topsis Score}. These tables show only the first five recommended foods for adult males and adult females.

\begin{table}[H]
\caption{Top 5 food choices for Person 2 (Female)}
\centering
\begin{tabular}{|c|c|c|}
\hline Serial No. & Food Name & Topsis Score \\
\hline 1 & Kebab and Butter naan & 0.874 \\
\hline 2 & Borhani & 0.694 \\
\hline 3 & Doughnuts & 0.645 \\
\hline 4 & Beef Nehari & 0.639 \\
\hline 5 & Swarma & 0.636 \\
\hline
\end{tabular}
\label{fr2}
\end{table}

\subsubsection{Result Analysis of Menu Planning}
Despite the fact that everyone's calorie intake differs, nutritionists feel that the average daily calorie intake at each meal should be divided down as follows: 300 to 400 calories for breakfast, 500 to 700 calories for lunch, and 500 to 700 calories for supper. Snacks should not have more than 200 calories. Snacks shouldn’t exceed 200 calories. For reference see  \href{https://globalnews.ca/news/3615212/this-is-what-your-breakfast-lunch-and-dinner-calories-actually-look-like/}{www.globalnews.ca} In total, intake calories should be between 1500 - 2000 calories. In order to avoid exceeding 2000 calories, we have planned our meal accordingly. Calories of the foods are measured on standard-sized plates. Tables \ref{menuplanning 1},\ref{menuplanning 2} represent a template of full-day menu planning for adult Male and Females. In Table \ref{menuplanning 1}, there is total of \textit{1676 Calories} for all the foods and in Table \ref{menuplanning 2} the total is \textit{1534 Calories}.

\begin{table}[H]
\caption{Menu Planning for Person 1 (Male)}
\centering
\begin{tabular}{| c || c| }
\hline Time & Food Name \\
\hline
\hline Breakfast & Bread and Butter (189 Cal), Omelete (154 Cal)  \\
\hline Lunch & Polao Roast (450 Cal)  \\
\hline Dinner & Kabab (691 Cal)  \\
\hline Snacks & Ramen (192.3 Cal) \\
\hline
\end{tabular}
\label{menuplanning 1}
\end{table}

\begin{table}[H]
\caption{Menu Planning for Person 2 (Female)}
\centering
\begin{tabular}{| c || c| }
\hline Time & Food Name \\
\hline
\hline Breakfast & Roti vegetable (388 Cal) \& Omelete (154 Cal)  \\
\hline Lunch & Rice with chicekn \& vegetable (450 Cal)  \\
\hline Dinner & Kacchi Biriyani (350 Cal)  \\
\hline Snacks & Chicken Shwarma (192 Cal) \\
\hline
\end{tabular}
\label{menuplanning 2}
\end{table}

\section{Conclusion}
Food recommendation and diet planning is challenging due to the diverse food choices and dietary requirements of individuals.  Unlike the traditional food recommendation system the proposed AMRP system considers users feelings, excitement, and preference affective states in meal recommendation. The affective states are predicted through EEG signals using hierarchical ensemble machine learning model. While the individual classification algorithms did not perform well, introducing  \textbf{voting classifier}, \textbf{hierarchical ensembling} had a great impact on the affective state classification. Due to higher accuracy of the models, food recommendation using TOPSIS becomes more promising. Sequentially, menu planning for a whole day has been more reliable and preference-centric.There is future scope to introduce deep learning based feature extraction method as well as Recurrent Neural Network (RNN) as affectivity classifier.

\bibliography{mybibfile}
\bibliographystyle{IEEEtran}


%





\ifCLASSOPTIONcaptionsoff
  \newpage
\fi

\end{document}